\documentstyle[amssymb,aps,prl,epsf,multicol]{revtex}

\begin{document}
\title{On the superconductivity in the system with preformed pairs}
\author{V. B. Geshkenbein}
\address{Theoretische Physik, ETH-H\"onggerberg, CH-8093, Z\"urich, \\
Switzerland \\
and Landau Institute for Theoretical Physics, Moscow, Russia}
\author{L. B. Ioffe}
\address{Physics Department, Rutgers University, Piscataway, NJ 08855, USA \\
and Landau Institute for Theoretical Physics, Moscow, Russia}
\author{A. I. Larkin}
\address{School of Physics and Astronomy, UMN-Minneapolis, MN 55455, USA \\
and Landau Institute for Theoretical Physics, Moscow, Russia }
\maketitle

\begin{abstract}
We discuss the phenomenology of the superconductivity resulting from the
bose condensation of the preformed pairs coexisting with unpaired fermions.
We show that this transition is more mean field like than usual bose
condensation, i.e. it is characterized by a relatively small value of the
Ginzburg parameter. We consider the Hall effect in the vortex flow regime
and in the fluctuational regime above $T_c$ and show that in this situation
it is much less than in the transition driven entirely by bose condesation
but much larger than in a usual superconductivity. We analyse the available
Hall data and conclude that this phenomenology describes reasonably well the
data in the underdoped materials of $YBaCuO$ family but is not an
appropriate description of optimally doped materials or underdoped $LaSrCuO$.
\end{abstract}



\section{Introduction}

It is known for a long time that excitation spectrum in underdoped high $%
T_{c}$ cuprates shows formation of the pseudogap at temperature $T_{s}$ far
above $T_{c}$; this phenomena was observed in the NMR responses \cite
{Takigawa91} and in optics \cite{Schlesinger,Basov96,Rice92}. Recently the
photoemission experiments showed that this phenomena can be attributed to
the electrons in the corners of the Fermi surface which acquire a gap in
these materials at about the same temperature at which pseudogap is observed
in optics and NMR \cite{Shen95}. Below $T_{c}$ the value of the gap does not
change significantly with temperature, instead the electron spectral
function develops coherence peaks at the gap edges. These data invite the
interpretation that this gap formation is due to the pairing of electrons in
the corners of the Fermi surface into the bosons which later bose condense
at $T_{c}$. The description of the superconductivity in the cuprates as a
bose condensation of preformed pairs was proposed also in different physical
contexts \cite{Aleksandrov93,Emery95,Ioffe96}. Unfortunately, all these
scenarios would lead to the conclusion that superconducting transition is
similar to the bose condensation and has a wide fluctuation region near $%
T_{c}$. This conclusion does not agree with the data which show that the
transition is more mean field like and that it is characterized by a small
value of Ginzburg parameter. In this paper we show that the bose
condensation description and mean-field nature of the superconducting
transition can be reconciled if bose condensation happens against the
background of the Fermi liquid and processes that convert bosons into the
fermions on the Fermi surface are allowed. We formulate the model which
describes this physics in Section II and derive its physical properties in
Section III. Another problem of the descriptions based on bose condensation
is that it leads to a large value of the Hall effect in the superconducting
state. Our analysis of the data shows that usual bose condensation is not
consistent with the data whereas bose condensation which happens against the
background of the Fermi liquid might be consistent with the available Hall
data in the underdoped $YBaCuO$ materials but is not consistent with the
data on optimally doped $YBaCuO$ or underdoped $LaSrCuO$. We emphasize
however that the data presentlly available are not sufficient to make the
definite conclusion, especially for the underdoped materials; we discuss the
data in more detail below in the Introduction and in Section IV. It is not
important for the foregoing discussion what is the microscopic mechanism
resulting in the formation of the preformed pairs but for the sake of
concreteness we shall discuss the model where these pairs are formed from
the electrons in the corners of the Fermi surface.

Qualitatively, the relative weakness of superconducting fluctuations in high
$T_c$ is clear from the following arguments. In these highly anisotropic
materials the coherence length in $c$-direction, $\xi_c(T=0)$, is  much
smaller than the interlayer distance, $d$ making them almost two dimensional
superconductors. In a purely two dimensional superconductor bose condensation
would show up as Berezinskii-Kosterlitz-Thouless transition in which  
the superfluid density jumps from $\rho _{S}(T_c) \approx \rho_S(0)$ to $\rho
_{S}=0$. Weak three dimensional effects would only smear this transition a
little.  Such $\rho_S(T)$ dependence was not observed in any cuprates; instead
the observed temperature dependence of $\rho _{S}(T)$ is  mean field like
in the broad temperature range even for underdoped cuprates, for instance in 
$YBa_{2}Cu_{4}O_{8}$ $\rho_{S}(T)\propto (T_{c}-T)$ for $T_{c}-T \gtrsim 
0.05T_{c}$ \cite{Schilling90}.
Note here that critical three dimensional behavior of optimally doped
$YBa_{2}Cu_{3}O_{7}$ reported in \cite{Hardy94} does not contradict the
conclusion that superconducting fluctuations are relatively weak. In this
material the $\rho_S(T)$ dependence remains linear in $T$ in a wide
temperature range \cite{Hardy93} and the mere fact that these critical
fluctuations are three dimensional implies that they occur only in the
vicinity of $T_c$ where the correlation length in $c$ direction becomes large,
$\xi_c \gg d$.

Quantitatively, the strength of superconducting fluctuations in quasi two
dimensional systems is determined by the superfluid density, $\rho_S$.
The measured absolute values of $\rho_S$ in cuprates turn out to be too large
for the bose condensation scenario; in $YBa_{2}Cu_{4}O_{8}$ the in-plane
penetration lengths are $\lambda _{a}=800\;\AA $ and $\lambda _{b}=2000\;\AA 
$ \cite{Basov95}. Such penetration length would lead to the
Kosterlitz-Thouless transition temperature $T_{KT}\approx 700\;K$ (here and
below we assume that the individual planes constituting bilayers are
strongly coupled so our estimates differ by a factor of 2 from the estimates
in \cite{Emery95}). This unrealistic value indicates that $\rho _{S}(T)$
must decrease by a factor of $10$ before the thermal fluctuations become
important in agreement with linear $\rho _{S}(T)$ dependence observed in 
\cite{Schilling90}. A related evidence of the weakness of superconducting
fluctuations is provided by a small value of the Ginzburg parameter which is
$G_i \sim 0.02$ in this material (see Eq. (\ref{Gi})).

Another important argument against bose condensation is provided by the Hall
effect data near $T_c$. Bose condensation of charged particles would lead
to a huge Hall effect in the superconducting state: $\sigma _{b}^{xy}=\frac{%
n_{b}ec}{B}$ where $n_{b}$ is density of bosons and a large fluctuational
contribution to the Hall effect above $T_c$. The existing data on the
underdoped materials show that the Hall effect in the flux flow regime is
large, but not as huge as follows from the bose condensation model.
Specifically, in $60\;K$ material we extrapolate the data obtained in the
flux flow regime at $T>15\;K$ \cite{Harris94} to zero temperature value $%
\sigma _{xy}=\frac{4\;10^{5}}{B[T]}\frac{1}{\Omega cm}$; this corresponds to
the effective boson density $n_{b}\approx 0.02$ per in-plane copper atom
which is too small.

A similar explanation of the pseudogap phenomena is based on the spin charge
separation model \cite{Ioffe96}. In this model the gap formation is due to
the pairing of spinons which carry no charge, such pairing does not lead to
superconductivity; it happens only at lower temperature and is due to bose
condensation of holons. This model has the same difficulty as the
condensation of the preformed pairs discussed above; there seems to be no
reason to expect a narrow fluctuation region if the transition is driven by
the bose condensation of holons.

A somewhat different view point on this problem is provided by the models
which interpolate between BCS like transition in Fermi liquid and ordinary
bose condensation of preformed pairs as the interaction strength is varied
\cite{DeMelo93,Levin96}. In this framework the data discussed above would make
one to conclude that high $T_c$ cuprates are well inside the Fermi liquid
regime and very far from the preformed pairs in contradiction to the observed
gap formation above $T_c$ in underdoped cuprates.

Another puzzling property of the superconducting transition is the change of
the Hall effect sign occurring below $T_{c}$. This sign change is preempted
by the negative fluctuational contribution to the positive Hall effect in
the normal state \cite{Lang95,Jin}; qualitatively both the sign change in
the superconducting phase and the fluctuational Hall effect can be explained
if Cooper pairs which are responsible for superconductivity are in fact
negatively charged. In this case these pairs give a large negative
contribution to the Hall conductivity in the vortex state below $T_{c\text{ }%
}$which is proportional to $1/B$; and produce negative fluctuational Hall
conductivity observed in \cite{Lang95}. Both Hall conductivity in the vortex
state near $T_{c}$ and the fluctuational conductivity above it can be
described in the framework of the time dependent Ginzburg Landau
(TDGL) equation;
for this equation the negative sign of the Cooper pair implies that the
imaginary part of the relaxation rate $%
\mathop{\rm Im}%
\gamma <0,$

\begin{equation}
\gamma \frac{\partial \Delta }{\partial t}=-\frac{\partial F}{\partial
\Delta ^{*}}  \label{TDGL}
\end{equation}
Here $F$ is the usual Ginzburg-Landau free energy \cite{Landau}.

In the framework of the usual BCS theory the sign and the magnitude of $%
\mathop{\rm Im}%
\gamma $ (and therefore the effective charge of the Cooper pair) is
determined by the derivative of the density of states at the Fermi surface, $%
\partial \nu /\partial \epsilon $, namely $%
\mathop{\rm Im}%
\gamma \sim -\partial \nu /\partial \epsilon $. This conclusion remains
valid for any weak coupling BCS\ theory of superconductivity regardless of
the nature of the interaction \cite{Aronov95}. For the high $T_{c}$ cuprates 
$\partial \nu /\partial \epsilon $ is controlled by the proximity to a van
Hove singularity and photoemission data show that Fermi surface is always a
hole like, so that $\partial \nu /\partial \epsilon <0$ and BCS theory would
predict the hole sign of $%
\mathop{\rm Im}%
\gamma $ in contrast to the data. One can also relate $\partial \nu
/\partial \epsilon $ to $\partial T_{c}/\partial \mu $ and avoid the use of
photoemission data, this would lead to the prediction $%
\mathop{\rm Im}%
\gamma \sim -\partial T_{c}/\partial \mu $ which implies that the
hydrodynamic contribution to the Hall effect is hole-like for the underdoped
cuprates and electron-like for the overdoped cuprates in a striking contrast
to the study of $La_{2-x}Sr_{x}CuO_{4}$ \cite{Matsuda96} which reported the
opposite correlation. We emphasize here that the sign change of the Hall
effect in the superconducting state does not itself contradicts the BCS
theory, it is only the disagreement between the sign of $\partial \nu
/\partial \epsilon $ (or $\partial T_{c}/\partial \mu $) and the sign of the
hydrodynamic contribution to the Hall effect which indicates that the weak
coupling BCS theory is not valid. In conventional, BCS-like superconductors,
the Hall effect might change sign if $-\partial \nu /\partial \epsilon $ has
the sign opposite to the sign of the charge carriers which is measured by
the normal state Hall effect. The sign of the charge carriers in the normal
state is determined by the topology of the Fermi surface. The sign change
might occur if $\partial \nu /\partial \epsilon <0$ on the electron-like
Fermi surface or if $\partial \nu /\partial \epsilon >0$ on the hole-like
Fermi surface.

Qualitatively, the notion of electron-like preformed pairs agrees with the
non-BCS behavior of the Hall effect of the superconductive pairs, but it is
difficult to reconcile both of them with the small value of the Ginzburg
parameter and with a moderate Hall effect in the superconducting state. In
this paper we resolve this dichotomy suggesting the model where preformed
pairs coexist with usual fermions and show that in such systems the Hall
effect might still be unusual but the Ginzburg parameter is small. One can
justify this model using the following qualitative arguments.

\centerline{\epsfxsize=6cm \epsfbox{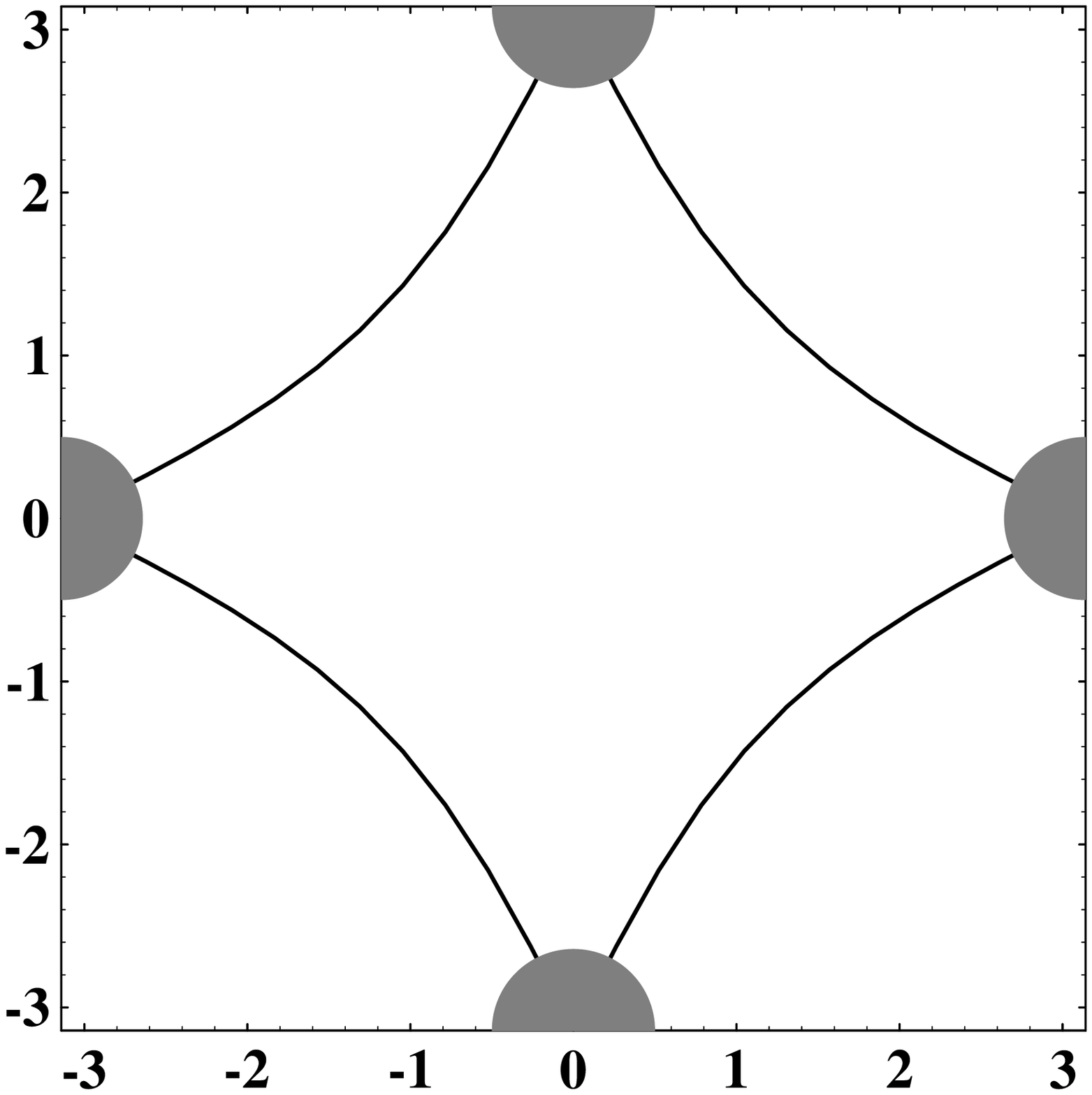}} 
{\footnotesize {\bf Fig 1. Sketch of the Fermi line and region of the
momentum space where pseudogap pairs is formed.} The Fermi line shown here
was obtained in the tight binding model with diagonal hopping $%
t^{\prime}=-0.3t$; it is similar to the Fermi line observed in the
underdoped $Bi_2Sr_2CaCu_2O_{8+\delta}$ \cite{Shen95}. The shaded discs
denote the part of the momentum space where a pseudogap was observed in the
experiment. We shall assume that the fermions in these regions are paired
into the bosons.} \vspace{0.5in} 

\section{Model}

It is well established that cuprate Fermi surface lies in the vicinity of
the van Hove points. Moreover, it is remarkable how small is dispersion of
fermions near $(\pi ,0)$ points in the underdoped cuprates according to the
photoemission data \cite{Shen95}. It is natural to assume that interaction
between these fermions can easily exceed their kinetic energy and that the
interaction with momentum transfer $q\sim (\pi ,\pi )$ is less repulsive
than interaction with small momentum transfer. Such interaction gives
fermions a gap which is due to the pairing in the antiferromagnetic or
superconducting channels. In the weak coupling approximation the d-wave
superconductive pairing dominates if the Fermi surface is not nested. In the
cuprates both photoemission data \cite{Shen95} and band structure
calculations \cite{Lichtenstein96} show that the Fermi surface is not
nested; a simplest Fermi surface which agrees with photoemission data shown
in Fig. 1. Therefore, it is reasonable to assume that fermions near the
corners of the Fermi surface (which lie inside the discs shown in Fig. 1)
are paired into bosons, $b^{\dagger }$, with charge $2e$ and no dispersion;
this is the key assumption of our model. So one-particle fermionic
excitations acquire a gap; the soft modes appearing instead of these
fermionic excitations are spinless bosons 
\begin{equation}
H=\varepsilon b_{q}^{\dagger }b_{q}  \label{H_b}
\end{equation}
where $\varepsilon $ is phenomenological parameter of the model. Note that
in this model the bose condensation does not occur because bosons have no
dispersion (i.e. are infinitely heavy). Another assumption of the model is
that interaction, $V$, transferring electrons from the 'discs', where they
are paired to the other parts of the Fermi surface (where Fermi velocity is
large) is weak. This assumption can be justified in the spinon-holon model
of charge separation \cite{Ioffe96} where this interaction is suppressed by
gauge field fluctuations. If $V$ is small we may neglect the effects of
these transfer processes on the gap formation in the corners of the Fermi
surface, clearly in this case the gap formation in the corners does not
necessarily result in the superconductivity and it does not give a gap to
the electrons away from the 'discs'. At higher temperatures the effects of
the remaining fermions can be neglected and bosons form a normal liquid
without long range order. Only at sufficiently low temperatures the
boson-mediated Cooper pairing between remaining unpaired electrons results
in the superconductivity.

Hamiltonian describing this physics is 
\begin{equation}
H=\sum_{q}\varepsilon b_{q}^{\dagger }b_{q}+\sum_{p,q}^{\prime
}V_{p,q}(b_{q}^{\dagger }c_{p\uparrow }c_{q-p\downarrow }+h.c.)+\sum_{p}\xi
_{p}c_{p,\sigma }^{\dagger }c_{p,\sigma }  \label{H}
\end{equation}
here $\sum^{\prime }$ denotes the sum over Brillouin zone excluding the
'disc' area.

Because $b$ describes fermions paired into the state with $d$-wave symmetry, 
$V_{p,q}$ also has this symmetry and we may approximate it by 
\begin{equation}
V_{p,q}=Va^2 (p_{x}^{2}-p_{y}^{2})  \label{V}
\end{equation}
and neglect its $q$-dependence at small $q$. Superconducting transition in
this model occurs at $T_{c}$ given by 
\begin{equation}
\epsilon =g \ln \frac{_{\Lambda }}{T_{c}},\quad g=\frac{1}{(2\pi )^{2}}\int
V_{p}^{2}\frac{dp}{v_{F}(p)}  \label{Tc}
\end{equation}
where integral $\int dp$ is taken over the Fermi line and $\Lambda \sim
\epsilon _{F\text{ }}$ is upper cut off.

Depending on the parameter $\epsilon$ model (\ref{H}) describes somewhat
different physical situations. At $\epsilon \gg T_c$ even at low
temperatures bosons exist only as virtual states, in this case the
superconducting transition is almost conventional. At $\epsilon \sim T_c$
the density of bosons at $T \sim T_c$ is significant so the superconducting
transition acquires some features of the bose condensation. We anticipate
that the former case is relevant for optimally doped cuprates whereas the
latter is more appropriate for the underdoped ones. Model (\ref{H}) is
somewhat similar to the model of disordered quasilocalized pairs coexisting
with Fermi liquid introduced in \cite{Eliashberg88}; in the latter model the
quasilocalized pairs are assumed to form resonances with energies $E$ that are
randomly distributed around the Fermi level. We do not know any experimental
justification for this assumption and we believe that the phase transition in
the presence of such large disorder in the energy levels will become quite
broad.

The superconducting transition at $T_{c}$ can be described as a bose
condensation which occurs only because bosons become coherent due to the
exchange of fermions. Alternatively, one might integrate out the bosons and
get the fermion model with retarded short-range interaction. Both approaches
lead to the same physical results. Here we shall adopt the bose formalism
because it is shorter and more physical in the regime when $\epsilon \sim T$
so the density of bosons is significant; we shall argue below that this
regime is relevant for the underdoped cuprates. At $T_c$ the gap begins to
open on the remaining part of the Fermi surface 
\begin{equation}
\Delta (p)=V(p)\langle b\rangle  \label{Delta}
\end{equation}
Clearly $\phi =\langle b\rangle $ plays the role of the order parameter in
this model; its thermal fluctuations are governed by the action $S(\phi )$
which is obtained after integrating out the fermion degrees of freedom: 
\begin{equation}
S(\phi )= \sum_\omega \phi^{*}_\omega \left\{ - i \gamma^{\prime \prime}
\omega - g\left[ \frac{T-T_{c}}{T_{c}} +\frac{\pi |\omega |}{8T_{c}} +\xi
_{0}^{2}\left| (\triangledown -\frac{2ie}{c}A) \right| ^{2} \right] \right\}
\phi_\omega - \frac{1}{2}\beta \int \left| \phi_t \right| ^{4} dt.
\label{S(phi)}
\end{equation}
Here we use imaginary time representation; we introduce coefficients 
\begin{equation}
\xi _{0}^{2}=\frac{7\varsigma (3)}{2(8\pi ^{2}T)^{2}g}\int
V_{p}^{2}v_{F}(p)dp,\quad \beta =\frac{7\varsigma (3)}{2(4\pi ^{2}T)^{2}}%
\int V_{p}^{4}\frac{dp}{v_{F}(p)}  \label{xi_0}
\end{equation}
and $\gamma^{\prime \prime} = - 1$, the latter we introduced to facilitate
the comparison with the usual time dependent Ginzburg-Landau equation where
this coefficient is determined by a particle-hole asymmetry near the Fermi
surface and is usually small. Generally, the coefficient $\gamma^{\prime
\prime}$ has contributions from the bare action of the bosons, $S_b =
- b^*\partial_t b - \epsilon b^*b$, described by Hamiltonian (\ref{H_b}) and
from the fermions that we integrated out but the latter is always small in
parameter $T_c/\epsilon_F$ leading to a simple result, $\gamma^{\prime
\prime}= - 1$.  

Action of a generic form (\ref{S(phi)}) but with
different parameter values describes also usual BCS type superconductivity in
the Fermi liquid, bose condesation and interpolation between these two
regimes \cite{DeMelo93}. The crucial difference between the interpolation
scheme \cite{DeMelo93} and model considered here is that the latter leads to
such parameters that the condesate amplitude, $|\phi|^2=g\tau/\beta$, always
remains small even far from $T_c$ and so the the fluctuation region is narrow
and the Hall effect never becomes too large.

\section{Results}

The gradient term in the effective action (\ref{S(phi)}) is determined by
the fermion properties. As a result the superconducting transition is mean
field like and thermal fluctuations become large only in the narrow
vicinity, $G_{i}$, of the transition temperature; it is convenient to
express it in terms of the screening length, $\lambda _{0}$ $G_{i}$ is given
by 
\begin{equation}
G_{i}=\frac{(4\pi \lambda _{0})^{2}T_{c}}{\sqrt{2}d\Phi _{0}^{2}}  \label{Gi}
\end{equation}
Here we define $\lambda _{0}$ as the value of the physical screening length
interpolated from the vicinity of the transition temperature to low
temperatures; it is expressed through the coefficients $g,\xi _{0}^{2}$ and $%
\beta $ of the effective action (\ref{S(phi)}) by 
\begin{equation}
\lambda _{0}^{2}=\frac{c^{2}\beta d}{32\pi e^{2}g^{2}\xi _{0}^{2}}
\label{lambda_0}
\end{equation}
We computed the coefficient $g,\xi _{0}^{2}$ and $\beta $ for the fermions
with the spectrum $\xi _{p}=-2t(\cos p_{x}+\cos py)-4t^{\prime }\cos
p_{x}\cos p_{y}-\mu $. Because fermions in the discs of size $p_{0}$ around
van Hove points are paired and do not contribute to the effective action we
excluded these regions from the integrals over the Fermi surface (\ref{xi_0}%
). We get 
\[
\lambda _{0}^{2}=\frac{c^{2}d}{e^{2}t}\Upsilon (\delta ),\quad G_{i}=\frac{16%
}{\sqrt{2}}\frac{T_c}{t}\Upsilon (\delta ). 
\]
Here $\Upsilon (\delta )$ is dimensionless function of the doping density $%
\delta $ which we plot in Fig. 2 for $t^{\prime }=-0.3t$ and different sizes
of the excluded regions, $p_{0}.$ We observe that once the regions near the
corners of the Fermi surface are excluded the doping dependence of the
penetration length becomes relatively weak and the dependance on the size of
the excluded regions become far more important. Qualitatively we expect that
the size of the excluded region becomes large in the underdoped bilayered
cuprates where large pseudogap was observed in spin responses and
photoemission so that their penetration length is larger than the one in the
optimally doped cuprates in agreement with the data. However we can not make
a quantitative comparison because we do not know the value of $p_0$.

\centerline{\epsfxsize=6cm \epsfbox{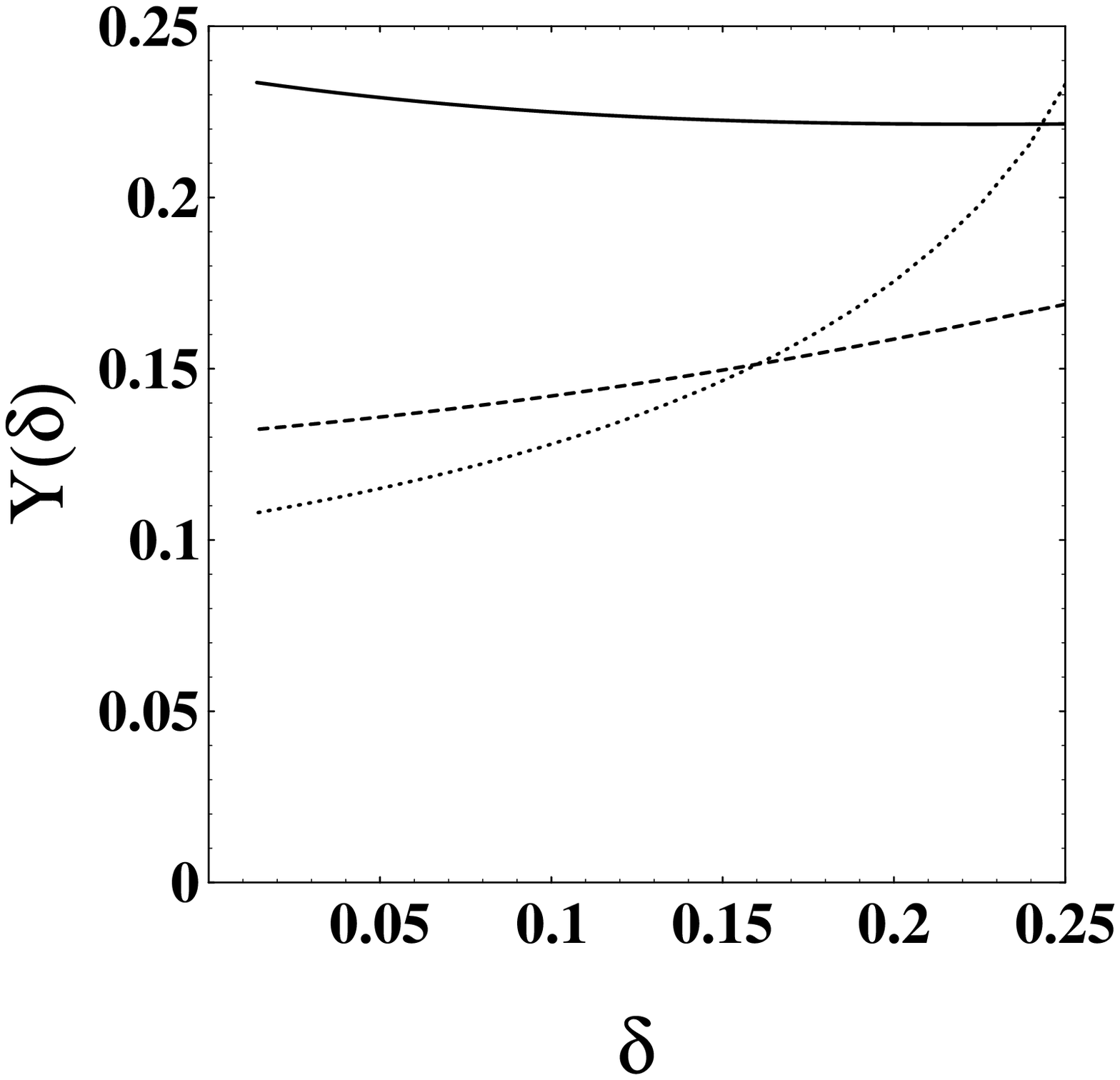}} 
{\footnotesize {\bf Fig 2. Dimensionless function $\Upsilon (\delta )$}
controlling the penetration depth $\lambda _{0}^{2}$ as a function of doping
assuming that other parameters are constants for different ``disc'' sizes $%
p_0$ where pseudogap is formed. Dotted line corresponds to $p_0=0$, dashed
line was obtained for $p_0=0.2 \pi$ and full line for $p_0=0.4 \pi$. } 
\vspace{0.5in} 

The results above do not depend on the dynamical part of the action (\ref
{S(phi)}), but it becomes important for the fluctuational conductivity \cite
{Aslamazov68}: 
\begin{equation}
\delta \sigma _{xx}=\frac{1}{16d}\frac{e^{2}}{
\rlap{\protect\rule[1.1ex]{.325em}{.1ex}}h%
}\frac{T_{c}}{T-T_{c}}  \label{delta_sigma_xx}
\end{equation}
where $d$ is the distance between planes. This result depends only weakly on
the properties of the electrons if they form a Fermi liquid even with a
large relaxation rate. Note that in the conventional bose condensation
scenario real part of the relaxation time is absent \cite{DeMelo93}
leading to a much larger fluctuational correction to the longitudinal
conductivity. 
Thus, it would be important to understand whether this universal behavior
(\ref{delta_sigma_xx}) is indeed observed in high $T_{c}$ cuprates.
Fluctuational Hall conductivity in low field is also controlled by the
coefficients of the effective action (\ref{S(phi)}) \cite{Aronov95,Ullah91}: 
\begin{equation}
\delta \sigma _{xy}=\frac{e^{2}}{3\pi d}\gamma ^{\prime \prime }\frac{ T%
}{g}\frac{eH\xi _{0}^{2}}{c
\rlap{\protect\rule[1.1ex]{.325em}{.1ex}}h%
}\left( \frac{T_{c}}{T-T_{c}}\right) ^{2}  \label{delta_sigma_xy}
\end{equation}
Here $\gamma ^{\prime \prime }$ is the coefficient of the non-dissipative
term in the action (\ref{S(phi)}), in this model $\gamma ^{\prime \prime
}=-1 $. This contribution should be added to the normal state Hall
conductivity. As a result a sign change of the Hall effect would occur above 
$T_{c}$ at 
\begin{equation}
\frac{T-T_{c}}{T}=\sqrt{2/3}\mid \frac{\gamma ^{\prime \prime } T_{c}}{g}%
\mid ^{1/2}\frac{\xi _{0}}{l},  \label{sign}
\end{equation}
where $l$ is the mean free path; here we used usual Drude formula, 
$\sigma_{xy}^n=\frac{nec}{B}(\omega_c\tau)^2$, for the
conductivity in the normal state. 
If $\mid \gamma ^{\prime \prime
} T_{c}/g\mid >(T_{c}/\mu )^{2}$ the correction to $\sigma _{xx}$ is
small and the Hall effect changes sign in the region where the longitudinal
conductivity is still close to the normal state value \cite{Jin}

In the vortex state the hydrodynamic contribution to the Hall effect is \cite
{Dorsey92,Kopnin93,FGLV,avo,LO} 
\begin{equation}
\sigma _{xy}^{V}=\frac{2ec}{B}\gamma ^{\prime
\prime }|\phi|^2=\frac{H_{c}^{2}(T)}{2\pi  (T_{c}-T)}\left( \frac{%
\gamma ^{\prime \prime } T_{c}}{g}\right) \frac{ec}{B}  \label{sigma^V}
\end{equation}
In the generic time dependent Ginzburg Landau theory \cite{Dorsey92,Kopnin93}
this contribution might be different by a numerical factor $\beta_V \simeq 1$;
the physical effect taken into account by this numerical factor is electric
field generated by the moving vortex. This effect is small and $\beta_V
\approx 1$ if the length, $\xi _{E}=4\xi_0\sqrt{2\sigma _{n}T_{c}\lambda ^{2}}
\approx \xi _{0} \sqrt{\frac{8}{\pi }\tau _{tr} T_{c}}$, which sets 
the scale for the electric field variations is long, $\xi _{E}\gg \xi _{0}$,
which seems to be an appropriate limit for cuprates. In conventional notations
\cite{Dorsey92,Kopnin93} TDGL dimensionless parameter $u=(\xi_E/\xi_0)^2 \ll 1$
for these materials.

In the bose condensation scenario $\gamma'' =1$, at low temperatures
$|\phi|^2$ coincides with the boson density and the Hall conductivity
$\sigma_{xy}=n_b ec/B$ is huge. In the present model $|\phi|^2 \sim
\frac{T^2}{g t} n_e$ is small leading to a smaller value of the Hall
conductivity. 

In the conventional BCS theory the coefficient $\gamma ^{\prime \prime }$ is
determined by the dependence of the density of states, $\nu (\mu )$ on the
chemical potential ($\gamma _{BCS}^{\prime \prime }=-\frac{g}{2}\frac{%
\partial \ln T_{c}}{\partial \mu }$); here it is controlled by the bosons
mediating the interaction between fermions. So, in the conventional BCS
theory $\left| \frac{\gamma ^{\prime \prime } T_{c}}{g}\right| $ is
small, $\left| \frac{\gamma ^{\prime \prime } T_{c}}{g}\right| \thicksim 
\frac{T_{c}}{\mu }$,  whereas here it is large. Formally we get a large $%
\left| \frac{\gamma ^{\prime \prime } T_{c}}{g}\right| $ in the model (%
\ref{H}) because we assumed that bosons are coupled to the pairs of
electrons, not holes which introduced a large particle-hole asymmetry. This
assumption can be justified if the fermion dispersion near van Hove points
is small so that properties of the bosons are determined by the relative
number of electrons and holes in the 'disc' area. Further, if the number of
electrons is small, the bosons are entirely electron-like and we get the
phenomenological model (\ref{H}); if the numbers of electrons and the holes
in the 'disc' area are close we would need to introduce two types of bosons
(electron-like and hole-like). This would lead to the effective action (\ref
{S(phi)}) with $\gamma ^{\prime \prime }\ll 1$ and the resulting Hall effect
would be much smaller.

These results show that $\gamma ^{\prime \prime }$ is not necessarily
related to $\frac{\partial \ln T_{c}}{\partial \mu }$ as was conjectured in 
\cite{Aronov95}. The arguments of \cite{Aronov95} were based on the gauge
invariance and on the assumption that $T_{c}$ dependence on $\mu $ implies a
dependence on the gauge invariant object $T_{c}(\mu +i\frac{d}{dt})$. This
is true in the BCS model with weak interaction where this dependence is due
to the density of states dependence on the chemical potential. However, in a
general case one should distinguish two sources of $T_{c}(\mu )$ dependence:
the dependence via the energy of pairing electrons, $\epsilon _{F}$ , and
the dependence via the total density of particles, $n$. The gauge invariance
indeed requires that $T_{c}(\epsilon _{F})$ is converted into the $%
T_{c}(\epsilon _{F}+i\omega )$ in the dynamical action but the
dependence via the total density is not modified by the frequency so
generally the quadratic term in the action is 
\begin{equation}
S^{(2)}=-\sum_{\omega }g\ln \left( \frac{T}{T_{c}(\epsilon _{F}+i\omega ,n)}%
\right) b_{\omega }^{*}b_{\omega }  \label{S^(2)}
\end{equation}
In other words, $n(\mu )$ dependence does not imply a non-gauge invariant
action, it can be reformulated in an explicitly gauge invariant manner as a
dependence on $\varphi =\nabla ^{-2}(\nabla E)$.

In the phenomenological model (\ref{H}) the $T_{c}$ dependence on the
doping, $\delta $, is due to the interaction term, $V(\delta )$ so that $%
T_{c}$ grows with doping. One possible microscopic mechanism of this
dependence is suppression of the interaction $V(\delta )$ by the gauge field
fluctuations discussed in \cite{Ioffe96} which becomes less in more doped
systems.

\section{Analysis of the data}

Equations (\ref{delta_sigma_xy},\ref{sigma^V}) can be directly compared with
the data. Note here that fluctuational Hall conductivity and Hall
conductivity in the flux flow regime are controlled by the same
dimensionless parameter $\left( \frac{\gamma ^{\prime \prime }T_{c}}{g}%
\right) $; this is a general feature of any hydrodynamic description based
on time-dependent Ginzburg-Landau equation. Experimental verification that
one gets the same parameter if it is extracted from the Hall data in the
fluctuational regime above $T_{c}$ and if it is extracted from the data in
the vortex flow regime would be a very important proof of the validity of
the hydrodynamic approach. The comparison of these parameters becomes more
complicated in weakly anisotropic materials such as $YBa_{2}Cu_{3}O_{7}$
where the fluctuational data are further complicated by the crossover
between two and three dimensional behaviors; to avoid these problems it is
better to compare the data obtained on more anisotropic materials. 

First we compare the values of $\left( \frac{\gamma ^{\prime \prime }T_{c}}{g}
\right) $ obtained on similar optimally doped materials. 
The extensive study \cite{Lang95} of the fluctuation regime in
$Bi_{2}Sr_{2}Ca_{2}Cu_{3}O_{x}$ shows that in the regime of 2D fluctuations
$\delta \sigma _{xy} \approx 0.08\;\frac{1}{\Omega cm}$ at $B=0.7\;T$. Using
the value $\frac{dH_{c2}}{dT} \approx 2\;T/K$ and $d=18.5\;\AA $ we obtain
$\left( \frac{\gamma ^{\prime \prime } T_{c}}{g}\right) \approx - 0.003$. 
Unfortunately we are not
aware of the Hall effect data in the vortex flow regime on this material, so
we compare this value with the other optimally doped cuprates. It is
convenient to characterize Hall conductivity data in the vortex flow regime by
the value of $\sigma_{xy}(0)$ obtained by a linear extrapolation to low
temperatures. For $YBa_{2}Cu_{3}O_{7}$ we use extrapolated value
$\sigma_{xy}(0)=\frac{2\;10^{5}}{B[T]}\frac{1}{\Omega cm}$ \cite{Harris95} and
$dH_{c}/dT=0.02\;T/K$; we get 
$\left(\frac{\gamma ^{\prime \prime } T_{c}}{g}\right) \approx -0.03$. For $%
Tl_{2}Ba_{2}CaCu_{2}O_{8}$ we use $\sigma_{xy}(0)=\frac{3\;10^{3}}{B[T]}
\frac{1}{\Omega cm}$ \cite{Samoilov94}, and $\frac{dH_c}{dT}=0.01\;K/T$ 
\cite{Wahl95,Samoilov96};
we get $\left(\frac{\gamma^{\prime\prime}T_{c}}{g}\right) \approx -0.0016$. 
The fit of the fluctuational  Hall conductivity data obtained on the same
sample agree with theoretical predictions if one chooses $dH_{c2}/dT=1\;T/K$
and  $\left(\frac{\gamma^{\prime \prime}T_{c}}{g}\right)\approx -0.002$
\cite{Samoilov96}.   These data indicate that hydrodynamic approach is likely
to be valid but do not allow to make a definite conclusion. They also show
that in optimally doped materials $\epsilon \sim g \gg T_{c}$, so the bosons
may exist only as virtual states of electron pair. 

The situation is different for underdoped bilayered cuprates. We take
extrapolated value $\sigma _{xy}=\frac{4\;10^{5}}{B}\frac{1}{\Omega cm}$ 
\cite{Harris94} and $dH_{c}/dT=0.006\;T/K$ \cite{Loram93} appropriate for
$60\;K$ $ YBa_{2}Cu_{3}O_{7-x}$; we get $\left( \frac{\gamma ^{\prime \prime
} T_{c}}{g}\right) \approx -0.9$ in agreement with our initial
expectations that bosons exist as real electron pairs in these materials.
However the data on the underdoped $La_{2-x}Sr_{x}CuO_{4}$ lead to a
different conclusion. Here we take $\sigma _{xy}=\frac{300}{B}\frac{1}{
\Omega cm}$ \cite{Matsuda96} and $dH_{c}/dT=0.006\;T/K$ \cite{Loram96} for the
material with $x=0.1$, we get $\left( \frac{\gamma ^{\prime \prime
} T_{c}}{g}\right) \approx -0.001$. This estimate implies that bosons
are unlikely exist as real pairs in underdoped $La_{2-x}Sr_{x}CuO_{4}$.
We emphasize that we do not know of any data which would allow us to check
that hydrodynamic approach remains valid for underdoped materials.

The independent check of the validity of the hydrodynamic (time dependent
Ginzburg-Landau) description in the flux flow regime is provided by the Hall
angle data {\it in weak field region $B\ll H_{c2}$}. In the framework
of the effective action (\ref{S(phi)}) it is
directly related with the same dimensionless parameter $\left(\frac{%
\gamma^{\prime \prime}  T_c} {g} \right)$ which we extracted from the
Hall conductivity \cite{Dorsey92,Kopnin93} 
\begin{equation}
\tan \theta_H = \frac{\gamma^{\prime \prime}}{\gamma^\prime \ln(\xi_E/\xi_0)}
= \left(\frac{\gamma^{\prime \prime}  T_c}{g} \right) \frac{8}{\pi
\ln(\xi_E/\xi_0)}  \label{tan_theta}
\end{equation}
The data \cite{Harris94} for Hall angle tangent in $60\;K$ $YBa_2Cu_3O_{7-x}$
show that its value extrapolated to $T=0$ is $\tan(\theta_H) \approx 1$, for 
$90\;K$ $YBa_2Cu_3O_{7}$ it is much smaller, $\tan(\theta_H) \approx 10^{-2}$%
, finally for $x=0.1$ $La_{2-x}Sr_xCuO_4$ $\tan(\theta_H) \approx 10^{-3}$ 
\cite{Matsuda96}. All these values are in a resonable agreement with the
above estimates for the parameter $\left(\frac{\gamma^{\prime \prime}  T_c%
}{g} \right)$ and usual expectation that $\ln(\xi_E/\xi_0) \sim 1$.

Another physical property of the phenomenological model (\ref{H}) is
anomalous thermopower in the normal state. The magnitude of this effect is
very sensitive to the value of $\epsilon_R /T$ where
$\epsilon_R=\epsilon-g\ln(\frac{\lambda}{T})$ is the effective chemical
potential of the pairs.  We have only a rough estimate 
of this parameter based on the following arguments. The boson density in the
phase space is $n_{0}=n_{B}(\epsilon_R /T)\lesssim 1$ (here $n_{B}$ is Bose
factor), so $\epsilon_R /T=\ln 1/n_{0}\gtrsim 1$; such $\epsilon_R $ makes
possible the scattering of electrons with energies larger than $\epsilon_R $
resulting in a large relaxation rate for these fermions. Because this
relaxation mechanism is effective only for fermions above the Fermi energy
it results in a large particle-hole asymmetry and leads to a large
thermopower. Assuming that this contribution to the relaxation rate $1/\tau
_{B}$ is much larger than the typical relation rate for the fermions with
energies less than $\epsilon_R $ we get Seebeck coefficient 
\begin{equation}
S_{0}=\ln (\frac{1}{n_{0}(T)})n_{0}(T)=\frac{g \ln(\frac{T}{T_c})}{T}
        \exp(-\frac{g \ln(\frac{T}{T_c})}{T}  \label{S_0}
\end{equation}
which is much larger than the usual value, $T/\epsilon _{F}$, for the normal
metal. The sign of the thermopower is positive. Its temperature dependence is
non-monotonic, at $G_i T_c \ll T-T_c \ll T_c$ the thermopower decreases with
temperature due to the temperature dependence of $\epsilon_R=g \ln(\frac{T}
{T_c})$, at higher temperatures, $T-T_c \gg T_c$ the temperature dependence of
$\epsilon_R$ becomes negligible and thermopower becomes small and it increases
with temperature. The sign and the value of the thermopower are in agreement
with the experiment \cite{Zhou96}, but its temperature dependence at high
temperatures is not. This is not very surprising because this model does not
describe the transport properties at high temperatures which are due to a new
physics associated with the appearance of low energy modes. The
disagreement between the predictions of the model (\ref{H}) and data implies
that these low energy modes are  responsible for the temperature dependence of
the thermopower at high temperatures.  

\section{Conclusion}

The model (\ref{H}) applies to the superconductivity in the underdoped
cuprates where gap opens above $T_{c}$, we expect a more usual transition in
the overdoped cuprates. The crossover from underdoped to overdoped occurs in
the framework of the model (\ref{H}) if $\epsilon$ and $V$ is increased with
doping; at large $\epsilon$ the transition can be described in terms of the
virtual pair formation and becomes very similar to a usual BCS picture.
However, even in this regime the contribution of these virtual pairs to the $%
\gamma^{\prime \prime }$ coefficient in time dependent Ginzburg Landau
equations can be much larger than the contribution coming from the density
of states dependence near the Fermi surface and may result in a sign change
of the Hall effect. In the optimally doped cuprates $n_{0}$ is still
non-zero and we expect a large hydrodynamic contribution to the Hall effect
and large positive thermopower.

In the optimally doped cuprates and in the underdoped ones above the
temperature of the pseudogap formation one expects new physical effects due
to the appearance of new low energy modes. These soft modes are responsible
for the anomalous transport relaxation rates. Another probe of the effect of
these modes in the optimally doped cuprates (where they are expected to
exist down to the transition temperature) is the fluctuational conductivity
which should no longer be given by universal form (\ref{delta_sigma_xx}). It
is important to determine experimentally whether fluctuational conductivity
agrees with a phenomenological Fermi liquid picture with large relaxation
rate which gives universal form (\ref{delta_sigma_xx}), if the data do not
fit the universal form (\ref{delta_sigma_xx}) it means that this
phenomenological Fermi liquid picture is not applicable at all even for the
in-plane properties.

A model similar to (3) but in real space also describes a phase transition
of the system of superconducting grains embedded in the normal matrix. In
this case the mixed $b^*cc$ term corresponds to the Andreev reflection at
the NS boundary. In this system the Hall effect in the superconducting state
is governed by the particle hole assymetry of the grains and may change sign
close to $T_c$.

In conclusion we have shown that the phenomenological description of the
superconductivity which follows from the concept of preformed pairs
coexisting with electrons on some patches of the Fermi surface agrees
semi-quantitatively with available data on Hall conductivity in the
fluctuation and flux flow regime and with the small value of the Ginzburg
parameter for underdoped bilayeres cuprates. However in order to describe the 
data on optimally doped bilayered cuprates or underdoped $LaSrCuO$ one needs
to assume that the value of the chemical potential for these pairs is large so
that preformed pairs exist only as virtual states. The important necessary
ingredients of this model are (1) the assumption that the pairs have very 
little dispersion of their own and (2) their coupling to the electrons on the
Fermi surface is weak. The hydrodynamic contribution to the Hall effect in
this model is controlled by the pairs and has electron-like sign; it
explains the Hall sign change observed experimentally.

It is not possible to test thoroghly the predictions of the model because
experiments which give data on the Hall and longitudinal conductivity in the
fluctuational and vortex flow regime obtained on the same sample are 
scarce. Such data on underdoped (spin gapped) materials do not exist at all.
It would be very important to verify experimentally that hydrodynamic
approach is still valid for the underdoped cuprates (i.e. that the parameters
extracted from the fluctuational regime are the same as those extracted from
vortex flow regime) and that the parameters needed for the hydrodynamic
description are indeed in agreement with the picture of preformed pairs
coexisting with fermions as we conclude here using a limited number of data.

We wish to thank G. Blatter, P. Coleman, A. Millis, and T. M. Rice for
helpful discussions and A. Samoilov for showing us his unpublished data. 
The major part of this work was performed at ITP, Santa
Barbara which hospitality and support via grant PHY94-07194 is gratefully
acknowledged; L. Ioffe is also grateful to ETH for its hospitality.

\end{document}